\documentstyle[11pt]{article}
\begin{document}

\centerline{\Large\bf A path-integral approach to the collisionless}
\vskip 5pt
\centerline{\Large\bf Boltzmann gas}
\vskip 30pt

\centerline{C. Y. Chen}
\centerline{Dept. of Physics, Beijing University of Aeronautics}
\centerline{and Astronautics, Beijing 100083, PRC}
\vskip 20pt
\centerline{Email: cychen@public2.east.net.cn}

\vfill

{\noindent{\bf Abstract}: 
On contrary to the customary thought, the well-known ``lemma'' that
the distribution function of a collisionless Boltzmann gas keeps invariant along
 a molecule's path represents not the strength but the weakness of 
the standard theory. One of its consequences states that the velocity  
distribution at any point is a condensed 
``image'' of all, complex and even discontinuous, structures of the 
entire spatial space.
Admitting the inability to describe the entire space with a microscopic
quantity, this paper introduces a new type of distribution 
function, called the solid-angle-average distribution function. With help of the
new distribution function, the dynamical behavior of collisionless Boltzmann gas
is formulated in terms of a set of integrals defined by molecular paths. 
In the new formalism, not only that the 
difficulties associated with the standard theory are 
surmounted  but also that some of practical 
 gases become calculable in terms of today's computer.    
}
\newpage
\section{Introduction}
The Boltzmann gas, by which we mean a gas in that molecules are subject to 
binary short-range interparticle forces, is usually studied by means of the  
Boltzmann equation\cite{reif}\cite{harris}. The conventional wisdom assumes  
that if a sufficiently large supercomputer were available, 
a finite-difference scheme could be employed to construct solutions of 
the Boltzmann equation 
and many realistic gases would become completely analyzable.
 
Some of difficulties related to the solution-construction scheme
of the Boltzmann equation 
are relatively well-known\cite{wirz}. They include the 
following. (i) Seven variables have to be dealt with (time, spatial 
coordinates and velocity components). (ii) The collisional operator in the 
Boltzmann equation is complex in nature. Due to these difficulties, only 
low-dimensional problems, for instance those having one dimension of 
coordinates and one dimension of velocity components, were tried in 
practical calculations.

In some of our works\cite{chen1}\cite{chen2}, it was argued that 
more fundamental difficulties existed with the standard kinetic theory. 
One of them was that distribution 
functions of Boltzmann gases involved in discontinuity
and quasi-discontinuity, thus the effectiveness of using a 
finite-difference scheme to solve the Boltzmann equation, as well as the 
validity of the Boltzmann equation itself, became questionable.

Putting aside whether the arguments presented by our previous works 
are sufficiently convincing or not, we will, in this paper, engage 
our primary attention 
with the following subject. Is there any alternative approach with the aid of 
that one can use today's computational means to calculate at least some of 
full-dimensional practical gases? A positive answer to this question will 
apparently arouse widespread interest.  

It turns out that if an average-type distribution function, called the 
solid-angle-average distribution function, is introduced
and the path-information of molecules is adequately taken into account, 
then the objective mentioned above can be accomplished.

The structure of this paper is the following. In Sec. 2, the standard 
approach to the Boltzmann gas is employed to ``derive'' a lemma that the 
distribution function of a collisionless Boltzmann gas, if continuous at the 
beginning, will keep its value along a molecule's path. In applying the 
lemma to various situations, we are led to an unexpected conclusion that 
strictly continuous and well-behaved distribution functions exist only on a 
theoretical assumption. In Sec 3, we examine two kinds of discontinuous 
distribution functions. The first kind is related to point-like molecular 
sources and the second kind to surface-like molecular sources. It is shown 
that the solid-angle-average distribution function needs to be 
introduced in order to treat discontinuous distribution functions related to
the molecular sources. 
Sec. 4 offers a comprehensive formulation in which both 
continuous and discontinuous distribution functions are formulated in terms 
of path-integrals defined by initial-state and boundary 
conditions. The final section, Sec. 5, gives a brief survey on essential 
features of the new formalism.    

\section{Evolution of continuous distribution function}

We suppose in this section that the distribution function of interest is 
perfectly continuous, differentiable in a more exact language, in the 
position-velocity space ($\mu$-space). 

Under this supposition, according to the standard kinetic theory, the 
collisionless Boltzmann equation holds and it 
reads 
\begin{equation}\label{bl} 
\frac{\partial f}{\partial t}+{\bf v}\cdot \frac{\partial f}{\partial {\bf r}} 
+\frac {{\bf F}}m \cdot\frac{\partial f}{\partial 
{\bf v}}=0, \end{equation}
in which the collisional effects between molecules have been disregarded (as 
the title of this paper has suggested). If the initial state 
\begin{equation}\label{ini}
 f(t_0+0,{\bf r},{\bf v})=f(t_0,{\bf r},{\bf v}) \end{equation}
and the boundary condition 
\begin{equation}\label{bou} f(t,{\bf r},{\bf v})=
\int K({\bf v},{\bf v}_1) f(t,{\bf r},{\bf v}_1)d{\bf v}_1, \end{equation}
where the shape of $K$ represents the collisional effects between molecules 
and boundaries\cite{kogan}, are regarded as known ones, the solution of Eq. 
(\ref{bl}) can be constructed by means of 
a finite-difference scheme\cite{wirz}. In such a scheme, notably, the governing 
equation (\ref{bl}) and the initial and boundary conditions 
(\ref{ini})-(\ref{bou}) constitute a complete equation set, and any 
additional information, such as the path-information of individual molecules, 
becomes theoretically dispensable. (In a sense, the whole idea behind 
introducing the conservation-type partial differential equations into kinetic 
theory is just to avoid knowing paths of individual molecules.) 

Instead of making a detailed analysis of the standard formalism outlined 
above, which has been done elsewhere\cite{chen1}, we now use the formalism 
to ``derive'' a relatively well-known lemma in statistical mechanics. The 
point here is that the lemma will automatically lead us to the very essence 
of gas dynamics.

By relating the vector variables ${\bf r}$ and ${\bf v}$ in (\ref{bl}) to a 
molecule's path, the partial differential equation can be rewritten in the 
form of a simple ordinary differential equation 
\begin{equation}\label{diff} 
\left.\frac{df}{dt}\right|_{{\bf r}(t),{\bf v}(t)}=0.\end{equation}
Of course, ${\bf r}(t)$ and ${\bf v}(t)$ are associated with the equations 
of motion for a single molecule
\begin{equation}\label{motion}   \frac{d{\bf r}(t)}{dt}=
{\bf v},\quad \frac{d{\bf v}(t)}{dt}=\frac{{\bf F}} m,\end{equation}
where $\bf F$ stands for all forces acting upon the molecule. 
Equation (\ref{diff}) simply means
\begin{equation}\label{fc} 
f(t,{\bf r},{\bf v})|_{{\bf r}(t),{\bf v}(t)}={\rm Constant}.\end{equation}
That is to say, in the absence of collisions a continuous distribution 
function keeps invariant along a molecular path in $\mu$-space.
[Since formula (\ref{fc}) is the integral of (\ref{diff}) along a path, we
may regard the formula as a primitive prototype of path-integral approach.]  
If the time development operator $T$ is employed to represent the equations 
of motion (\ref{motion}), namely if we have
\begin{equation} 
T(t_0,t)[{\bf r}(t_0),{\bf v}(t_0)]=[{\bf r}(t),{\bf v}(t)],\end{equation}
then we can equivalently express (\ref{fc}) as 
\begin{equation}\label{ft} f(t,{\bf r},{\bf v})=
f(t_0,{\bf r}_0,{\bf v}_0)=f(t_0,T^{-1}{\bf r},T^{-1}{\bf v}),\end{equation}
where ${\bf r}_0={\bf r}(t_0)$, ${\bf v}_0={\bf v}(t_0)$ and $T^{-1}=T(t,t_0)$. 
In view of (\ref{ft}) and (\ref{fc}), we will refer to $f(t_0)$  as the 
``source'' function and refer to $f(t)$ as the ``image'' function. In a similar 
spirit, it will be said that ${\bf r}_0$ is a source point and ${\bf r}$ an image 
point.

Now, we use the path-invariance expressed by (\ref{fc}) to derive some
dynamical features of the Boltzmann gas. These features, though quite 
 interesting and very essential, received no enough attention before. 
For simplicity only situations in that no external force exists will be 
considered, though  equations (\ref{diff})-(\ref{ft}) are valid more 
generally. 

Suppose, to begin with, that the initial distribution function 
$f(t_0,{\bf r},{\bf v})$ is known (which is continuous) and we are concerned with the 
velocity distribution at a specific point ${\bf r}$, denoted 
by $f_{\bf r}(t,{\bf v})$ hereafter. Equation (\ref{ft}) shows that
\begin{equation} f_{\bf r}(t,{\bf v})=f(t_0,T^{-1}{\bf r},{\bf v}).\end{equation}
For one fixed value of $v_i=|{\bf v}_i|$, the source points ${\bf r}_i=T^{-1}{\bf r}$ 
simply form
a spherical surface, labeled as $S_i=S(v_i)$ in Fig. 1. The radius of such a 
surface can be expressed by 
\begin{equation} \label{sph} |{\bf r}_i-{\bf r}|= v_i \cdot (t-t_0).\end{equation}
To determine the velocity distribution, we must let $v_i$ take values from very 
small to very large, and thus the radius $|{\bf r}-{\bf r}_i|$ varies from very 
small to very large.
This implies that the velocity distribution $f_{\bf r}(t,{\bf v})$ 
is a condensed image of the entire spatial space. If $f(t_0,{\bf r},{\bf v})$ is not 
spatially 
uniform, $f_{\bf r}(t,{\bf v})$ will possess a complex structure, infinitely 
complex in general situations (which reminds us of the famous Cantor set and 
many other 
interesting structures in the fractal studies). To make the situation even 
more worrisome, the inference presented above also suggests that any 
discontinuous things, if exist in the spatial space, will constantly create 
discontinuous images on velocity distributions in the nearby and distant 
regions (virtually everywhere). 

To see the point in another perspective, let's assume that a numerical work
is being carried out and the velocity space is being divided into many small but 
finite cells (for time being pay less attention to the spatial space), of 
which one takes the form $\Delta{\bf v}=  v^2 \Delta v\Delta \Omega$.
It is then up to the investigator to determine the distribution function
for each of the cells, 
denoted by 
\begin{equation} f_{\bf r}(t,\Delta v, \Delta \Omega). \end{equation}
In other words, 
each of $f_{\bf r}(t,\Delta v, \Delta \Omega)$ should be endowed with
one definite value.
The task, though seems quite straightforward, cannot be done conventionally.
Look at Fig. 2, in which the spatial cone $-\Delta \Omega$ 
corresponding to the velocity cone $\Delta \Omega$, called the {\it 
effective cone} hereafter, and the two 
spherical surfaces $S_i=S(v_i)$ and $S_{i+1}=S(v_{i+1})$, where $v_{i+1}-v_i=\Delta v$,
are plotted.
It is easy to find that when ${\bf v}$ varies within the  
range $v^2\Delta v \Delta \Omega$, the source point $T^{-1}{\bf r}$ runs over the 
shaded volume in the figure. Since 
the thickness of the shaded volume is proportional to $\Delta v$ solely (for 
the fixed $t-t_0$), if we let $\Delta v$ be smaller and smaller the 
value of 
\begin{equation}\label{limit1} \lim\limits_{\Delta v\rightarrow 0}
 f_{\bf r}(t,\Delta v, \Omega) \end{equation}
tends to a limit in the usual sense. Whereas, the size of $\Delta S_i$, the area element 
on $S(v_i)$ enclosed by the effective cone, is proportional to $v_i^2$, and if 
we let $\Delta \Omega$ be smaller and smaller the value of  
\begin{equation}\label{limit2} \lim\limits_{\Delta \Omega\rightarrow 0} 
f_{\bf r}(t,v,\Delta \Omega) \end{equation}
tends to a limit at a rate that strongly depends on the magnitude of $v$. 
That is to say, 
the velocity distribution at ${\bf r}$ is not uniformly continuous.
According to mathematics textbooks\cite{rudin}\cite{gel}, a function that is not 
uniformly continuous may involve some irregularities and has to be treated carefully.
As far as our situation is concerned, there indeed exist
difficult things.  As one thing, when $v_i$ is rather large, the area of 
$\Delta S_i$ must be rather large. If the source function $f(t_0)$ on this large area
element varies significantly (quite possible in practical situations), 
giving a unique value to 
the image function $f_{\bf r}(t,\Delta v, \Delta\Omega)$ becomes a tricky business.
This type of difficulty has been named as {\it quasi-discontinuity} in 
our previous works\cite{chen1}\cite{chen2}.

We now study boundary effects. At first suppose, 
if the path-invariance expressed by (\ref{fc}) is combined with the boundary 
condition (\ref{bou}), the boundary effects can be regarded as being  
formulated completely. The investigation below, however, tells a different story. 
First of all, note that a realistic boundary cannot be considered as having a 
geometrically and physically uniform surface. This, according to the spirit 
of calculus, leaves us no choice but to divide the boundary into many  
area elements, infinitesimally small in the theoretical sense.
We are then supposed to examine how each of them receive and emit incident 
molecules. If we assume that  an area element $d S$ in Fig. 3 uniformly reemit
the incident molecules and the distribution function immediately above it is known as 
$f({\bf r}_o,{\bf v})$, it follows from (\ref{fc}) that, assuming the gas to be
relatively stationary, 
\begin{equation}\label{ff}
f({\bf r}_o,{\bf v})= f({\bf r}_{p_1},{\bf v})= f({\bf r}_{p_2},{\bf v}).\end{equation}
 An interesting question must arise. What is 
the difference between the velocity distributions at $p_1$ and at $p_2$? Fig. 
3b offers manifestation that the velocity distribution at $p_1$ takes the value 
of $f({\bf r}_o,{\bf v})$ within the solid-angle range $d \Omega_1$ while the velocity
distribution at $p_2$ takes the same value within $d \Omega_2$. An essential fact is
that the infinitesimally small $d\Omega_1$ is significantly larger than 
the infinitesimally small $d\Omega_2$. It is then rather obvious that
equation (\ref{ff}) alone cannot be deemed as a satisfactory description of the process.
(The next section shows that introducing 
$\delta$-functions into the scheme does not provide much help either.)

We have seen that the path-invariance lemma, though intended to 
describe the evolution of continuous distribution function, 
discloses many features that cannot be well treated in terms of continuous 
distribution functions. 

Before finishing this section and turning our attention to the next subject,
we wish to comment on the differences between the Boltzmann-equation approach and 
the approach represented by formulas (\ref{diff})-(\ref{ft}), though it
seems that formula (\ref{diff}) or (\ref{fc}) is ``derived from''
and ``completely equivalent to'' the collisionless
Boltzmann equation (\ref{bl}).

According to most textbooks, the derivation of the Boltzmann equation is
closely related to the conservation law in the phase space $\mu$-space\cite{reif}.
Ref. 4 and Ref. 5, however, argue that such a
conservation law is not truly sound. On one hand, 
defining fluxes through five-dimensional hypersurfaces (six of them can enclose
a six-dimensional volume element) is an absolute must for the conservation law; 
on the other hand, such fluxes cannot be well defined in either physical or mathematical
sense. 
For instance, in defining a flux through 
the ordinary surface $dydz$ we invoke that the velocity
$\dot x=v_x$ is perpendicular to the surface; whereas in defining a flux through the
hypersurface $dv_x dv_y dv_z dydz$ we cannot invoke that the 
velocity $\dot x=v_x$ is perpendicular to the hypersurface, because
 $v_x$ itself is among the five dimensions of the hypersurface.

Formula (\ref{diff}) can be derived independently through a 
Jacobian approach under  two assumptions\cite{reif}\cite{harris}. 
One is that the distribution function
is perfectly continuous at an initial time. 
The other is that forces acting on all molecules are
free from dissipation (independent of speed) 
and free from fluctuation (smooth everywhere). 
After obtaining (\ref{diff}), equation (\ref{bl}) can be obtained 
by abandoning the path-information expressed by equation
(\ref{motion}). Noting that the abandonment involves an information loss,
we wish to say that the approach represented by 
(\ref{diff})-(\ref{ft}), instead of the collisionless Boltzmann equation (\ref{bl}), 
should be regarded as a more basic and more complete formalism.
The discussion of this section, in which
expressions (\ref{diff})-(\ref{ft}), instead of (\ref{bl}), reveal the
essential features of continuous distribution function, also substantiates 
the statement presented above.  

The standard kinetic theory gets heritages from the 
ordinary fluid mechanics. It assumes that a local value of distribution 
function is largely influenced by its immediate neighborhood and the 
influence is exerted through two-dimensional surfaces or 
five-dimensional hypersurfaces.
By adopting this assumption, the theory unanimously  
employs a conservation-type partial differential equation to set up its 
framework. Formulas (\ref{diff})-(\ref{ft}), however, suggest  
that events of gas dynamics develop along molecular paths, 
which are one-dimensional lines in $\mu$-space. 
The new picture, quite different from the standard one in the physical and 
mathematical senses, implies that 
even a gas that initially possesses a 
perfectly continuous distribution function will not behave itself like
a continuous medium. We have manifestly seen this point
 in the discussion related to the quasi-discontinuity.

Finally, formulas (\ref{diff})-(\ref{ft}) offer, 
in a primitive manner though, different concepts 
concerning the way we approach 
to gas dynamics. Instead of relying on comprehensive, but difficult and  
delicate, differential equations,  one is virtually led to accepting a 
procedure-type approach. Such approach supposedly
includes the following steps: (i) to determine paths of 
individual molecules; (ii) to formulate how a local distribution function 
gives a contribution to another local distribution function through a path; 
(iii) to integrate all contributions associated with all possible molecular 
paths. The last step is necessary since, according to the basic principles 
of statistical mechanics, molecular paths must involve a probabilistic 
nature due to the molecule-boundary and molecule-molecule interaction.  
(All points in this paper are given in terms of classical mechanics.) 
We will, by following Feynman\cite{feynman}, 
call an approach of this type the path-integral approach.

\section{Evolution of discontinuous distribution function}

The last section has shown that the 
 Boltzmann gas should be regarded less as a perfectly 
continuous medium and more as a special collection of individual molecules.
In this section we will formulate how discontinuous distribution functions,
each of which describes a set of molecules, develop along
molecular paths.

For simplicity only discontinuous distribution 
functions that are produced by point-like sources and surface-like sources
are of our interest. It is also assumed that molecules of the interested gases 
are free from external forces. If needed, the approach here can, in a 
straightforward way, be adapted to more general situations. 

In Fig. 4, a point ${\bf r}_0 $ is plotted at which molecules are generated 
(for whatever reasons). To make the discussion applicable later, the 
molecular emission rate $\rho$ is allowed to depend on the time, velocity 
magnitude and velocity direction. The number of molecular emission can thus 
be expressed by
\begin{equation}\label{rho} \rho(t_0 ,{\bf r}_0 ,v,\Omega_0 )
 dt_0  dv d\Omega_0 , \end{equation}
where $\Omega_0$ is defined in the frame whose origin is at 
${\bf r}_0$. Moving along a molecule's path, it is found that at a later 
time $t$ the emitted molecules spread over the volume 
\begin{equation}\label{volume}  |{\bf r}-{\bf r}_0 |^2 v dt d\Omega_0 \end{equation}
where ${\bf r}$ is the point in the spatial space such that 
\begin{equation} {\bf r}-{\bf r}_0 ={\bf v} (t-t_0 ). \end{equation}
It follows from (\ref{rho}) and (\ref{volume}) that the density of molecules 
at ${\bf r}$ is 
\begin{equation} n({\bf r})= \frac{ \rho dv dt_0  d\Omega_0 }
{|{\bf r}-{\bf r}_0 |^2 v dt d\Omega_0 }
= \frac{ \rho dv }{|{\bf r}-{\bf r}_0 |^2 v},\end{equation}
where $dt=dt_0 $ has been understood. Since the distribution function 
$f$ at the point ${\bf r}$ satisfies
\begin{equation} \int_\Omega f v^2 dv d\Omega=n({\bf r})
 =\frac{\rho dv}{|{\bf r}-{\bf r}_0 |^2 v},\end{equation}
the distribution function produced by the point-like source is 
\begin{equation}\label{ps} f(t,{\bf r},v,\Omega)=\frac {\rho(t_0 ,{\bf r}_0 
,v,\Omega_0 )}
{|{\bf r}-{\bf r}_0 |^2 v^3}\delta(\Omega-\Omega_0 ). \end{equation}
Note that $\Omega_0 $ in (\ref{ps}) takes the same direction as that of 
$({\bf r}-{\bf r}_0 )$. The form of (\ref{ps}) manifests that molecules produced by
a point-like source are associated with a discontinuous distribution function.
If one tries to apply the Boltzmann equation (\ref{bl}) to it, difficulties arise
sharply with the differentiation operations.

Now, we consider a surface-like source of molecules. It seems that after the 
surface is divided into many small area elements and the formula Eq. 
(\ref{ps}) is applied to each of them, our task is virtually accomplished. However, the 
following discussion shows that if we stick to what the standard kinetic 
theory implies, more troublesome things will emerge and no progress can be made.  
  
Let $\Delta S_{0i}$ denote one of area elements on the surface and 
$\eta(t_0, {\bf r}_0, v, \Omega_0)$ denote the molecular emission rate 
per unit area on $\Delta S_{0i}$. By identifying 
\begin{equation}\label{sigma} \eta(t_0,{\bf r}_0,v,\Omega_0)
 \Delta S_{0i} \end{equation}
with the molecular emission rate $\rho$ in (\ref{ps}), we find the entire 
distribution function produced by the surface-like source to be 
\begin{equation}\label{fsur} f(t,{\bf r},v,\Omega)=\sum\limits_i\frac  {\eta \Delta 
S_{0i}} {|{\bf r}-{\bf r}_{0i}|^2 v^3}\delta(\Omega-\Omega_{0i}), \end{equation}
where $i$ runs over all the elements on the surface. If one rewrites 
(\ref{fsur}) in the integral form
\begin{equation}\label{int} f(t,{\bf r},v,\Omega)=\int \frac  {\eta d 
S_0}{|{\bf r}-{\bf r}_0|^2 v^3}
\delta(\Omega-\Omega_0),\end{equation}
the following question will be of the immediate concern.  Does equation 
(\ref{int}) define a normally behaved distribution function? Fig. 5 
demonstrates that each of $\Omega_0$ points to a different direction 
while the distribution function on the left side of (\ref{int}) involves 
only one direction defined by $\Omega$. That is to say, expression 
(\ref{int}) cannot be integrated in the usual sense.

It is quite interesting to look at the peculiarities that we have just 
encountered. In dealing with our discontinuous distribution functions
 neither the usual differentiation nor the 
usual integration works smoothly. 

In order to find a way out of the difficult situation, we propose to use the 
following {\it solid-angle-average} distribution function such that if the 
exact distribution function $f(t,{\bf r},{\bf v})$ is known, 
which exists only in a pure academic sense, then  
the average distribution function is
\begin{equation}\label{aver} \bar f(t,{\bf r},v,\Delta \Omega)= \frac 1{\Delta 
\Omega} \int_{\Delta \Omega} f(t,{\bf r},{\bf v}) d\Omega,\end{equation}
where $\Delta\Omega$ is a solid-angle range in the velocity space set by 
the investigator. In practical calculations, it is convenient to employ the 
spherical coordinate system of velocity, in which $\Omega$ is defined by the 
polar angle $\theta$ and the azimuthal angle $\phi$, and the solid angle 
range can be expressed by
\begin{equation} \Delta \Omega \approx \sin \theta \Delta\theta \Delta \phi.
\end{equation}
On this understanding, the entire solid-angle range of velocity 
associated with a spatial point 
${\bf r}$ is divided into a large number of small, but finite, ranges, which may 
or may not be equal to each other. Of course, if the distribution function 
is smooth enough and $\Delta \Omega$ is sufficiently small, we may simply 
assume \begin{equation} \bar f(t,{\bf r},v,\Delta\Omega) = f(t,{\bf r},v,\Omega). 
\end{equation}               
In the rest of this paper, we will always omit the bar notation when 
referring to such distribution function.

In terms of practical calculations, the size of $\Delta \Omega$ should be 
chosen properly so that the computational work can be done efficiently and 
at the same time no significant macroscopic phenomena will be overlooked. 

As we may notice, an investigator of numerical work can take a similar 
strategy to treat distribution functions in terms of $\Delta{\bf r}$ and $\Delta 
v$. The difference here lies in that our solid-angle-average distribution 
function is introduced in a theoretical (analytical) consideration: the 
discontinuity represented by (\ref{ps}) and the quasi-discontinuity 
represented by (\ref{limit2}) should, and have to,
 be handled under the new definition.

It is rather important to emphasize that the distribution function 
introduced above seems to be an ``approximate'' one, but it actually 
represents an ``accurate'' approach in the following two senses. One is that 
by letting $\Delta \Omega$ be sufficiently small, we can describe a 
statistical process with any desirable accuracy.  The other is that errors 
related to giving up the exact distribution function 
are largely inherent to nature not to the way we approach to it. 

With help of the new average distribution function, the deterministic nature 
and the probabilistic nature of a statistical process can be kept in a 
balanced way. By using all kinds of mathematical operations, such as  
differentiation and integration, to formulate the distribution function, 
we preserve the macroscopic causality. By taking the solid-angle average, 
some of the microscopic information, in particular those related to 
discontinuity and quasi-discontinuity, are forsaken forever.

We are now equipped to formulate the discontinuous distribution function due 
to the existence of a surface-like source, as shown in Fig. 6. For a chosen 
solid-angle range $\Delta\Omega$ in the velocity space there is an effective 
cone $-\Delta \Omega$ in the spatial space. A surface-like molecular source 
enclosed by the effective cone, the shaded area $\Delta S$ in Fig. 6, gives 
contributions to $f(t,{\bf r},v,\Delta\Omega)$.

Assuming the emission rate on the surface to be known and allowing expression
(\ref{int}), not normally behaved though, to represent the emitted 
molecules, we obtain from (\ref{aver}) 
\begin{equation}\label{fsur2} f=\frac 1{\Delta\Omega}\int\int_{\Delta S}
 \frac  {\eta d S_0}{|{\bf r}-{\bf r}_0|^2 v^3}
\delta(\Omega-\Omega_0) d\Omega. \end{equation}
By exchanging the order of the integration, we finally arrive at    
\begin{equation}\label{fsur3} f(t,{\bf r},v,\Delta\Omega)=\frac 1{\Delta\Omega} 
\int_{\Delta S}  \frac  {\eta(t_0,{\bf r}_0,v,\Omega_0) d S_0 
}{|{\bf r}-{\bf r}_0|^2 v^3}.\end{equation}
In the integrand of (\ref{fsur3}) ${\bf r}_0$ is the position of 
$dS_0$, $t_0$ is equal to $t-|{\bf r}-{\bf r}_0|/v$ and $\Omega_0$ 
points to the direction of $({\bf r}-{\bf r}_0)$.

It is easy to see that the distribution function expressed by (\ref{fsur3}) 
is finite and well behaved. 

\section{The complete path-integral formulation and its application}
The discussion in the last two sections has shown that the distribution 
function at any specific point consists of two parts. The first part is a 
continuous one produced by the continuous distribution function existing 
previously; the second part is a discontinuous one produced by surface-like 
molecular sources. Here, we summarize the last two sections and give a 
comprehensive formulation for the dynamics of collisionless Boltzmann gas.

In Fig. 7 we assume that the complete solid-angle-average distribution 
function $f(t,{\bf r},v,\Delta\Omega)$ is affected by a continuous distribution 
function and a surface-like source. Both the ``sources'' exist within the 
effective cone and constantly ``emit'' molecules into the region around the 
point ${\bf r}$. The complete distribution function can formally be expressed as
\begin{equation}\label{f} f(t,{\bf r},v,\Delta\Omega)=f_{(i)}+f_{(ii)},\end{equation}
where $f_{(i)}$ and $f_{(ii)}$ are 
produced by the aforementioned two sources  respectively.

First, we wish to determine $f_{(i)}$ in terms of certain specifications of 
initial state. Providing the initial state of the continuous distribution 
function, denoted by $f^{ct}$,
 is given only at one specific moment $t_0$, we have no other 
choices but to take the following approach.
All relevant source points, according to Sec. 2, can 
be determined by the mapping $T^{-1}{\bf r}$ and by the effective cone $-\Delta 
\Omega$. Namely, they distribute on an area element of the spherical surface
$|{\bf r}_0-{\bf r}|= v(t-t_0)$, denoted by $\Delta S_1$ in Fig. 7.
 We will refer to such area element as a virtual effective surface 
in view of that similar surface-like sources (boundaries)
 are physical ones. From (\ref{fc}) 
and (\ref{aver}), the continuous part of the  distribution function can be 
written as an integral 
\begin{equation}\label{final1}\begin{array}{lll}
f_{(i)}(t,{\bf r},v,\Delta\Omega)
\!\!&\!\!=\!\!&\!\! \displaystyle{ \frac 1{\Delta\Omega}  \int_{\Delta 
 \Omega} f^{ct}(t,{\bf r},v,\Omega)  d\Omega } \vspace{4pt}\\
\!\!&\!\!=\!\!&\!\! \displaystyle{ \frac 1{|{\bf r}-{\bf r}_0|^2 \Delta\Omega} 
\int_{\Delta S_1} f^{ct}(t_0, {\bf r}_0,v,\Omega_0) 
U_{{\bf r}_0{\bf r}} dS_0} \end{array}\end{equation}
where ${\bf r}_0$ is the position of $d S_0$, $\Omega_0$ is in the 
direction $({\bf r}-{\bf r}_0)$, and $U_{{\bf r}_0{\bf r}}$ is a specially defined 
path-clearness function such that 
\begin{equation}\label{path} U_{{\bf r}_0{\bf r}}=\left\{ \begin{array}{ll} 1& {\rm no\; 
block\; along \; the\; path \;from\;} {{\bf r}_0}\;{\rm to}\;{\bf r}\\
0 & {\rm otherwise}. \end{array}\right.\end{equation}
Note that the difficulty of quasi-discontinuity discussed in Sec. 2 is no 
longer an issue: $f_{(i)}$ is uniquely defined and makes 
appropriate physical sense no matter how dynamically 
the source function $f^{ct}(t_0)$ varies on the virtual effective 
surface $\Delta S_1$.

If the distribution function is known within a period of time (before $t$), 
we have freedom to choose the virtual effective surface. For instance, by 
choosing the surface $\Delta S_1^\prime$ in Fig. 7, we have, as an alternative of 
(\ref{final1}), 
\begin{equation}\label{final2}
f_{(i)}= \frac 1{\Delta\Omega}  \int_{\Delta S_1^\prime}\frac{ 
f^{ct}(t_0,{\bf r}_0,v,\Omega_0) 
U_{{\bf r}_0{\bf r}}}{|{\bf r}-{\bf r}_0|^2}|\cos\alpha| dS_0 ,\end{equation}
where $\alpha$ is the angle between the direction $({\bf r}-{\bf r}_0)$ and the 
normal of the area element $dS_0$, and the time $t_0$ now depends 
on the position of $dS_0$, namely $t_0 =t-|{\bf r}-{\bf r}_0|/v$.
  
Then, we wish to formulate the discontinuous part of the distribution 
function. As suggested in the last two sections, boundary surfaces in a gas 
can be treated as surface-like molecular sources, since they constantly 
reemit incident molecules. In Fig. 7, the boundary surface $\Delta S_2$ 
within the effective cone, called the physical effective surface, is singled 
out as a surface-like source that can affect $f(t,{\bf r},v,\Delta\Omega)$.

The focus is naturally on the local emission rate 
$\eta(t_0,{\bf r}_0,v,\Omega_0)$ defined by (\ref{sigma}). To 
determine it, we first consider the falling rate of incident molecules 
expressed by  
\begin{equation}\xi(t_0,{\bf r}_0,v_i,\Omega_i) =f(t_0,{\bf r}_0,v_i,\Omega_i)
  \cos\theta_i U(\cos\theta_i) ,\end{equation}  
where $\theta_i$ is the angle between the inward normal of the local surface 
and the incident direction of the molecules and $U$ is a step function 
whose value is equal to unity if $\cos\theta_i>0$ and equal to zero otherwise. 
 
As well known, the functional relation between $\eta$ and $\xi$ has 
to be ultimately measured in experiments\cite{kogan}, though some kinds of 
theoretical models may be of use at some stages. To see how this relation 
can be formulated empirically, consider a molecule that moves with the velocity 
$(v_i,\Omega_i)$, strikes an area element and leaves the area 
element with a velocity within the range $dv d\Omega_0$, with respect 
to the surface element, in a certain 
probability. If the probability is denoted by
\begin{equation} P(v_i,\Omega_i,v,\Omega_0)dv d\Omega_0\end{equation}
and its normalization takes the form
\begin{equation}\label{nor} \int P(v_i,\Omega_i,v,\Omega_0)dv d\Omega_0=1,
\end{equation}
then the emission rate of the area element, for rarefied gases, will 
satisfy 
\begin{equation}\label{sigma1} \eta(t_0,{\bf r}_0,v,\Omega_0)=\int
P(v_i,\Omega_i,v,\Omega_0) \xi v_i^2dv_id\Omega_i.\end{equation} 
To verify the 
correctness of (\ref{sigma1}), one may integrate (\ref{sigma1}) and obtain, 
with the help of  (\ref{nor}), 
\begin{equation} dt_0 dS_0 
\int \eta(t_0, {\bf r}_0, v, \Omega_0) dv d\Omega_0=
 dt_0 dS_0\int \xi(t_0,{\bf r}_0,v_i,\Omega_i) v_i^2 dv_i d\Omega_i,\end{equation}
which is nothing but the conservation law of the molecular number on the 
surface element.
It is now clear that the purpose of such experiment should be to determine 
the functional form of $P(v_i,\Omega_i,v,\Omega_0)$. 
  
Under the assumption that the molecular emission rate $\eta$ of the 
involved boundary surface has been determined (by whatever means), the interested 
discontinuous part can, by virtue of Eq. (\ref{fsur3}), be written as
\begin{equation}\label{final3} f_{(ii)}=\frac 1{\Delta\Omega} \int_{\Delta S_2}\frac 
{\eta(t_0, {\bf r}_0,v,\Omega_0) dS_0}{|{\bf r}-{\bf r}_0|^2 
v^3}U_{{\bf r}_0{\bf r}},   \end{equation}
where the path-clearness function $U_{{\bf r}_0{\bf r}}$ has been defined by 
(\ref{path}). 

The final result is then
\begin{equation}\label{finalf}\begin{array}{r}
\displaystyle{ f(t,{\bf r},v,\Delta \Omega) =
 \frac 1{\Delta\Omega}  \int_{\Delta S_1^\prime}\frac{ 
f^{ct}(t_0,{\bf r}_0,v,\Omega_0)|\cos\alpha|} 
{|{\bf r}-{\bf r}_0|^2}U_{{\bf r}_0{\bf r}} dS_0 \qquad} \vspace{4pt} \\
\displaystyle{  +\frac 1{\Delta\Omega} \int_{\Delta S_2}\frac 
{\eta(t_0, {\bf r}_0,v,\Omega_0) }{|{\bf r}-{\bf r}_0|^2 
v^3}U_{{\bf r}_0{\bf r}} dS_0  ,} \end{array}  \end{equation}
where the first term of the right side can be replaced by (\ref{final1}).
If macroscopic quantities are of interest, the distribution function 
expressed by (\ref{finalf}) can be used as a conventional one.

At this point, it is interesting to comment on the time-irreversibility of 
the formalism given above. 
It is kind of well-known that the time-reversibility dilemma related to a 
dynamical process can be eliminated if a proper statistical average is taken. 
With the statistical average, two 
issues get involved. (i) Some pieces of microscopic information are forsaken. (ii) 
Conservative basic forces are converted into fluctuating and dissipative 
(speed-dependent) forces. Notably, the second issue mentioned above  
is directly responsible for the time-irreversibility of the interested process. 
In this approach, it appears that the time-irreversibility has nicely been embedded.
Though the pure evolution of continuous distribution function 
expressed by (\ref{fc}) or (\ref{ft}) is time-reversible, other
treatments, including the definition of the solid-angle-average 
distribution function and the formulation of boundary effects,  are
manifestly time-irreversible.
In particular, by allowing the molecule-boundary interaction to be determined by
empirical laws (or other adequate statistical laws), the formulation in this paper 
is fully in harmony with the Langevin theory\cite{pathria}, in which
fluctuating and dissipative 
forces naturally arise from the interaction between a moving body and its surrounding
molecules.

At the end of this section, we give a brief look at what will happen if  
a calculation suggested by this paper is practically performed. 
Fig. 8 offers schematic of a gas 
leaking out of a large container through a small hole. 
(Ref. 4 shows that the standard theory 
encounters many difficulties in treating the case.) Since the container is 
rather large one may assume that the distribution function on a 
surface, labeled as $S$ in the figure, constantly takes the value 
\begin{equation}\label{max} f=
n_0 \left(\frac m{2\pi\kappa T}\right)^{3/2} 
\exp\left(-\frac{mv^2}{2\kappa T}\right),\end{equation}
provided that the surface $S$ is not very close to the hole. It is obvious 
that the values of distribution function at the starting points of paths 1,2 and 3 
are all equal to the value of (\ref{max}). If contributions from complex 
paths, such as path 3, are neglected, we can directly use an ordinary 
PC-type computer to calculate (\ref{finalf}) on the 
understanding that the molecular emission rate $\eta$ in the formulation 
has been determined empirically.

\section{Conclusion}

We have set up a new formalism that exhibits many features strongly 
different from those related to the standard theory. They can briefly be 
listed as the following. 

\begin{enumerate}
\item
Instead of pursuing the exact distribution function, a special type of 
distribution function, called the solid-angle-average distribution function, 
is introduced. The new-type distribution function can describe statistical 
phenomena with any desirable accuracy.

\item 
Instead of relying on a partial differential equation, which is difficult 
and delicate, the new approach gives a set of tamable integral formulas. As 
well known, for a law of nature the integral formulation often enjoys 
advantages. 
\item Instead of disregarding the path-information, the new approach bases 
its formulation on the path-information of molecules. With the 
path-information included, many sophisticated and important features of 
kinetic systems become explicable and treatable.

\item
Instead of treating the collisionless Boltzmann gas as a completely 
continuous medium, the new approach treats the discontinuity and 
quasi-discontinuity at the very beginning. No singularity of any 
type exists with the final result of the formalism.

\item 
Instead of falling into the time-reversibility paradox, the new approach 
admits its inability to formulate all microscopic details. In treating 
boundary effects, the time irreversibility has explicitly been embedded in the 
formalism. \end{enumerate}

The complete dynamics of the Boltzmann gas, including collisional effects, 
will be formulated in different works\cite{chen2}\cite{chen3}.
Helpful discussion with Prof. Keying Guan is greatly appreciated. This work 
is partly supported by the fund provided by Education Ministry, PRC.

\newpage
\centerline{\bf\Large Figure captions}

\begin{enumerate}
\item
 Local distribution  as an image of the entire space.
\item 
 Source points affecting the local distribution within 
$v^2 \Delta \Omega \Delta v$.
\item
 Local distributions near an area element.
\item
 A point-like molecular source.
\item
 Schematic of discontinuous distribution produced by a surface-like source.
\item
 A surface-like molecular source within an effective cone. 
\item
 One physical surface and two virtual surfaces within an effective cone.
\item
 Molecules leaking out of a large container.

\end{enumerate}

\newpage

\noindent {\bf Figure 1}

\setlength{\unitlength}{0.013in} 
\begin{picture}(200,155)

\put(112,65){\makebox(35,8)[l]{\scriptsize $S_1$}}
\put(162,57){\makebox(35,8)[l]{\scriptsize  $S_2$}}
\put(160,114){\makebox(35,8)[l]{\scriptsize  $S_i$}}
\put(125,83){\makebox(35,8)[l]{\scriptsize  $\bf r$}}
\put(58,104){\makebox(35,8)[l]{$y$}}
\put(175,15){\makebox(35,8)[l]{$x$}}
\put(50,20){\line(1,0){120}}
\put(171,20.2){\vector(1,0){1}}
\put(60.6,100){\vector(0,1){1}}
\put(60,15){\line(0,1){85}}

\put(130,80){\line(1,1){32}}
\put(130,80){\line(1,2){20}}
\put(130,80){\line(-2,1){22}}
\put(130,80){\line(-1,0){25}}
\put(130,80){\line(1,-2){16}}
\put(130,80){\line(4,-3){27}}

\put(150,100){\vector(-1,-1){1}}
\put(143.0,107){\vector(-1,-2){1}}
\put(116.2,85.80){\vector(2,-1){1}}
\put(115,80.5){\vector(1,0){1}}
\put(138.5,61){\vector(-1,2){1}}
\put(145,68){\vector(-4,3){1}}

\put(155.00,80.00){\circle*{1}}
\put(155.00,80.00){\circle*{1}}
\put(154.76,83.48){\circle*{1}}
\put(154.03,86.89){\circle*{1}}
\put(152.84,90.17){\circle*{1}}
\put(151.20,93.25){\circle*{1}}
\put(149.15,96.07){\circle*{1}}
\put(146.73,98.58){\circle*{1}}
\put(143.98,100.73){\circle*{1}}
\put(140.96,102.47){\circle*{1}}
\put(137.73,103.78){\circle*{1}}
\put(134.34,104.62){\circle*{1}}
\put(130.87,104.98){\circle*{1}}
\put(127.39,104.86){\circle*{1}}
\put(123.95,104.26){\circle*{1}}
\put(120.63,103.18){\circle*{1}}
\put(117.50,101.65){\circle*{1}}
\put(114.61,99.70){\circle*{1}}
\put(112.02,97.37){\circle*{1}}
\put(109.77,94.69){\circle*{1}}
\put(107.93,91.74){\circle*{1}}
\put(106.51,88.55){\circle*{1}}
\put(105.55,85.20){\circle*{1}}
\put(105.06,81.74){\circle*{1}}
\put(105.06,78.26){\circle*{1}}
\put(105.55,74.80){\circle*{1}}
\put(106.51,71.45){\circle*{1}}
\put(107.93,68.26){\circle*{1}}
\put(109.77,65.31){\circle*{1}}
\put(112.02,62.63){\circle*{1}}
\put(114.61,60.30){\circle*{1}}
\put(117.50,58.35){\circle*{1}}
\put(120.63,56.82){\circle*{1}}
\put(123.95,55.74){\circle*{1}}
\put(127.39,55.14){\circle*{1}}
\put(130.87,55.02){\circle*{1}}
\put(134.34,55.38){\circle*{1}}
\put(137.73,56.22){\circle*{1}}
\put(140.96,57.53){\circle*{1}}
\put(143.98,59.27){\circle*{1}}
\put(146.73,61.42){\circle*{1}}
\put(149.15,63.93){\circle*{1}}
\put(151.20,66.75){\circle*{1}}
\put(152.84,69.83){\circle*{1}}
\put(154.03,73.11){\circle*{1}}
\put(154.76,76.52){\circle*{1}}
\put(165.00,80.00){\circle*{1}}
\put(165.00,80.00){\circle*{1}}
\put(164.83,83.48){\circle*{1}}
\put(164.31,86.94){\circle*{1}}
\put(163.45,90.32){\circle*{1}}
\put(162.25,93.60){\circle*{1}}
\put(160.74,96.74){\circle*{1}}
\put(158.92,99.72){\circle*{1}}
\put(156.81,102.50){\circle*{1}}
\put(154.44,105.06){\circle*{1}}
\put(151.82,107.36){\circle*{1}}
\put(148.99,109.40){\circle*{1}}
\put(145.97,111.15){\circle*{1}}
\put(142.79,112.58){\circle*{1}}
\put(139.48,113.69){\circle*{1}}
\put(136.08,114.47){\circle*{1}}
\put(132.62,114.90){\circle*{1}}
\put(129.13,114.99){\circle*{1}}
\put(125.65,114.73){\circle*{1}}
\put(122.21,114.12){\circle*{1}}
\put(118.85,113.18){\circle*{1}}
\put(115.60,111.90){\circle*{1}}
\put(112.50,110.31){\circle*{1}}
\put(109.57,108.42){\circle*{1}}
\put(106.84,106.24){\circle*{1}}
\put(104.34,103.81){\circle*{1}}
\put(102.10,101.13){\circle*{1}}
\put(100.13,98.25){\circle*{1}}
\put(98.47,95.19){\circle*{1}}
\put(97.11,91.97){\circle*{1}}
\put(96.08,88.64){\circle*{1}}
\put(95.39,85.22){\circle*{1}}
\put(95.04,81.74){\circle*{1}}
\put(95.04,78.26){\circle*{1}}
\put(95.39,74.78){\circle*{1}}
\put(96.08,71.36){\circle*{1}}
\put(97.11,68.03){\circle*{1}}
\put(98.47,64.81){\circle*{1}}
\put(100.13,61.75){\circle*{1}}
\put(102.10,58.87){\circle*{1}}
\put(104.34,56.19){\circle*{1}}
\put(106.84,53.76){\circle*{1}}
\put(109.57,51.58){\circle*{1}}
\put(112.50,49.69){\circle*{1}}
\put(115.60,48.10){\circle*{1}}
\put(118.85,46.82){\circle*{1}}
\put(122.21,45.88){\circle*{1}}
\put(125.65,45.27){\circle*{1}}
\put(129.13,45.01){\circle*{1}}
\put(132.62,45.10){\circle*{1}}
\put(136.08,45.53){\circle*{1}}
\put(139.48,46.31){\circle*{1}}
\put(142.79,47.42){\circle*{1}}
\put(145.97,48.85){\circle*{1}}
\put(148.99,50.60){\circle*{1}}
\put(151.82,52.64){\circle*{1}}
\put(154.44,54.94){\circle*{1}}
\put(156.81,57.50){\circle*{1}}
\put(158.92,60.28){\circle*{1}}
\put(160.74,63.26){\circle*{1}}
\put(162.25,66.40){\circle*{1}}
\put(163.45,69.68){\circle*{1}}
\put(164.31,73.06){\circle*{1}}
\put(164.83,76.52){\circle*{1}}

\put(172.85,93.74){\circle*{1}}
\put(171.66,97.02){\circle*{1}}
\put(170.21,100.20){\circle*{1}}
\put(168.53,103.25){\circle*{1}}
\put(166.61,106.17){\circle*{1}}
\put(164.47,108.93){\circle*{1}}
\put(162.13,111.51){\circle*{1}}
\put(159.59,113.90){\circle*{1}}
\put(156.87,116.10){\circle*{1}}
\put(153.99,118.07){\circle*{1}}
\put(150.97,119.81){\circle*{1}}
\put(147.82,121.32){\circle*{1}}
\put(144.57,122.58){\circle*{1}}
\end{picture}

\vskip 20pt
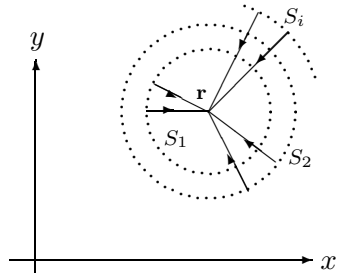
\noindent {\bf Figure 2}

\setlength{\unitlength}{0.013in} 

\begin{picture}(200,155)
\put(58,105){\makebox(35,8)[l]{$y$}}
\put(174,15){\makebox(35,8)[l]{$x$}}
\put(50,20){\line(1,0){120}}
\put(170,20.4){\vector(1,0){1}}
\put(60.6,100){\vector(0,1){1}}
\put(60,15){\line(0,1){85}}

\put(110,70){\line(3,2){37}}
\put(110,70){\line(1,2){20}}
\put(108,60){\makebox(35,8)[l]{$\bf r$}}
\put(120,55){\makebox(35,8)[l]{\scriptsize  $-\Delta \Omega$}}
\put(116,78){\line(1,-1){14}}

\put(137,73){\makebox(35,8)[l]{\scriptsize $S_i$}}
\put(152,78){\makebox(35,8)[l]{\scriptsize  $S_{i+1}$}}

\put(128,105){\line(1,0){9}}
\put(128,101){\line(1,0){14}}
\put(133,97){\line(1,0){11}}
\put(136,93){\line(1,0){7}}

\put(151.66,87.02){\circle*{1}}
\put(150.21,90.20){\circle*{1}}
\put(148.53,93.25){\circle*{1}}
\put(146.61,96.17){\circle*{1}}
\put(144.47,98.93){\circle*{1}}
\put(142.13,101.51){\circle*{1}}
\put(139.59,103.90){\circle*{1}}
\put(136.87,106.10){\circle*{1}}
\put(133.99,108.07){\circle*{1}}
\put(130.97,109.81){\circle*{1}}
\put(127.82,111.32){\circle*{1}}
\put(124.57,112.58){\circle*{1}}

\put(142.25,83.60){\circle*{1}}
\put(140.74,86.74){\circle*{1}}
\put(138.92,89.72){\circle*{1}}
\put(136.81,92.50){\circle*{1}}
\put(134.44,95.06){\circle*{1}}
\put(131.82,97.36){\circle*{1}}
\put(128.99,99.40){\circle*{1}}
\put(125.97,101.15){\circle*{1}}
\put(122.79,102.58){\circle*{1}}
\put(119.48,103.69){\circle*{1}}
\end{picture}

\vskip 20pt
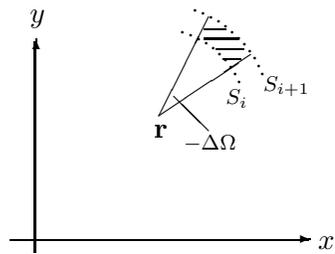
\noindent {\bf Figure 3}

\setlength{\unitlength}{0.013in} 

\begin{picture}(200,155)
\put(70,0){\makebox(35,8)[l]{\bf (a)}}
\put(170,0){\makebox(35,8)[l]{\bf (b)}}
\multiput(80,40)(100,0){2}{\oval(30,20)}

\put(120,80){\circle*{3}}
\put(150,110){\circle*{3}}

\multiput(230,90)(30,30){2}{\line(1,-2){8}}
\put(258,93){\makebox(35,8)[l]{\scriptsize $ d\Omega_2$}}
\put(228,63){\makebox(35,8)[l]{\scriptsize $ d\Omega_1$}}
\multiput(107,87)(96,0){2}{\makebox(35,8)[l]{$p_1$}}
\multiput(137,113)(100,0){2}{\makebox(35,8)[l]{$p_2$}}

\put(200,35){\makebox(35,8)[l]{\scriptsize $d S$}}
\put(98,35){\makebox(35,8)[l]{\scriptsize $d S$}}
\put(85,45){\line(1,1){65}}
\put(85,45){\circle*{3}}
\multiput(80,30)(100,0){2}{\line(0,1){6}}
\multiput(85,30)(100,0){2}{\line(0,1){6}}
\multiput(90,32)(100,0){2}{\line(0,1){6}}
\multiput(75,30)(100,0){2}{\line(0,1){6}}
\multiput(70,32)(100,0){2}{\line(0,1){6}}

\multiput(220,80)(-3,-1.8){16}{\circle*{1}}
\multiput(220,80)(-1.8,-3){14}{\circle*{1}}
\put(220,80){\line(5,3){20}}
\put(220,80){\line(3,5){12}}
\multiput(246.8,107.6)(-3.2,-2.4){25}{\circle*{1}}
\multiput(246.8,107.6)(-2.4,-3.2){22}{\circle*{1}}
\put(246.8,107.6){\line(4,3){20}}
\put(246.8,107.6){\line(3,4){15}}
\put(220,80){\circle*{3}}
\put(246.8,107.6){\circle*{3}}
\end{picture}

\newpage
\noindent {\bf Figure 4}

\setlength{\unitlength}{0.013in} 
\begin{picture}(200,155)
\put(58,105){\makebox(35,8)[l]{$y$}}
\put(174,15){\makebox(35,8)[l]{$x$}} 
\put(50,20){\line(1,0){120}}
\put(170,20.2){\vector(1,0){1}}
\put(60.6,100){\vector(0,1){1}}
\put(60,15){\line(0,1){85}}
\put(120,80){\vector(1,0){50}}
\put(120,82){\vector(4,1){50}}
\put(120,78){\vector(4,-1){50}}
\put(112,80){\circle*{3}}
\put(177,80){\circle*{3}}
\put(108,85){\makebox(35,8)[l]{${\bf r}_0$}} 
\put(174,83){\makebox(35,8)[l]{$\bf r$}} 
\put(205,75){\makebox(35,8)[l]{\scriptsize $\Omega_0$}} 
\put(185,80){\vector(1,0){13}}
\end{picture}

\vskip 20pt
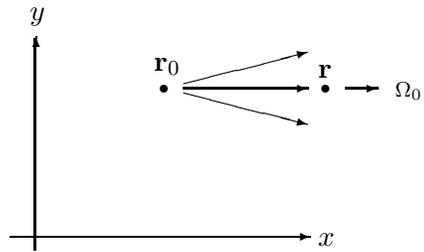
\noindent {\bf Figure 5}

\setlength{\unitlength}{0.013in} 
\begin{picture}(200,155)
\put(58,105){\makebox(35,8)[l]{$y$}}
\put(174,15){\makebox(35,8)[l]{$x$}} 
\put(50,20){\line(1,0){120}}
\put(170,20.2){\vector(1,0){1}}
\put(60.6,100){\vector(0,1){1}}
\put(60,15){\line(0,1){85}}
\put(136,78){\makebox(10,8)[l]{$\bf r$}}
\put(110,85){\makebox(10,8)[l]{\scriptsize $dS_{01}$}}
\put(110,56){\makebox(10,8)[l]{\scriptsize $dS_{02}$}}
\put(170,80){\makebox(10,8)[l]{\scriptsize $\Omega_{02}$}}
\put(170,60){\makebox(10,8)[l]{\scriptsize $\Omega_{01}$}}

\put(107,67){\oval(5,8){}}
\put(107,83){\oval(5,8){}}
\put(100,75){\oval(20,30){}}
\put(90,75){\line(1,0){8}}
\put(90,70){\line(1,0){8}}
\put(90,80){\line(1,0){8}}
\put(91,85){\line(1,0){8}}
\put(91,65){\line(1,0){8}}

\put(107,67){\vector(4,1){60}}
\put(107,83){\vector(4,-1){60}}
\put(139,75){\circle *{4}}
\end{picture}

\vskip 20pt
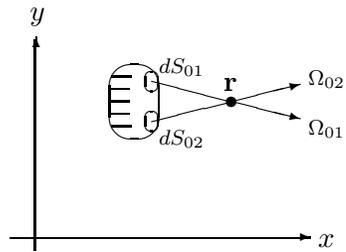
\noindent {\bf Figure 6}

\setlength{\unitlength}{0.013in} 
\begin{picture}(200,155)
\put(58,105){\makebox(35,8)[l]{$y$}}
\put(174,15){\makebox(35,8)[l]{$x$}} 
\put(50,20){\line(1,0){120}}
\put(170,20.4){\vector(1,0){1}}
\put(60.6,100){\vector(0,1){1}}
\put(60,15){\line(0,1){85}}

\put(120,102){\makebox(10,8)[l]{\scriptsize $\Delta S$}}
\put(170,79){\makebox(10,8)[l]{\scriptsize $\Delta\Omega$}}
\put(140,57){\makebox(10,8)[l]{\scriptsize $-\Delta\Omega$}}
\put(138,83){\line(1,-1){15}}
\put(110,85){\line(1,1){15}}
\put(107,83){\oval(6,20){}}
\put(100,83){\oval(20,30){}}
\multiput(104.5,74)(0,3){7}{\line(1,0){5}}
\put(151,83){\circle*{4}}
\put(148,86){\makebox(10,8)[l]{$\bf r$}}
\put(167,87){\line(-4,-1){60}}
\put(167,79){\line(-4,1){60}}
\end{picture}

\newpage
\noindent {\bf Figure 7}

\setlength{\unitlength}{0.013in} 
\begin{picture}(200,155)
\put(58,105){\makebox(35,8)[l]{$y$}}
\put(174,15){\makebox(35,8)[l]{$x$}} 
\put(50,20){\line(1,0){120}}
\put(170,20.2){\vector(1,0){1}}
\put(60.6,100){\vector(0,1){1}}
\put(60,15){\line(0,1){85}}
\put(117,46){\makebox(10,8)[l]{\scriptsize$\Delta S_2$}}
\put(105,110){\makebox(10,8)[l]{\scriptsize $\Delta S_1$}}
\put(200,79){\makebox(10,8)[l]{\scriptsize $\Delta\Omega$}}
\put(178,86){\makebox(10,8)[l]{$\bf r$}}
\put(149,105){\makebox(10,8)[l]{\scriptsize $\Delta S_1^\prime$}}
\put(110,72){\line(1,-1){15}}
\put(126,87){\line(1,1){22}}
\put(86,96){\line(1,1){18}}

\put(100,68){\oval(20,30){}}
\multiput(85,83)(4,0){5}{\circle*{1}}
\put(181,83){\circle*{4}}
\put(197,87){\line(-4,-1){88}}
\put(197,79){\line(-4,1){110}}

\put(85.00,83.00){\circle*{1}}
\put(85.06,86.35){\circle*{1}}
\put(85.23,89.70){\circle*{1}}
\put(85.53,93.03){\circle*{1}}
\put(85.93,96.36){\circle*{1}}
\put(86.46,99.67){\circle*{1}}
\put(87.10,102.96){\circle*{1}}
\put(87.85,106.22){\circle*{1}}

\put(120, 96.7){\circle*{1}}
\put(121.3, 94.3){\circle*{1}}
\put(122.7, 91.7){\circle*{1}}
\put(124.2, 89.2){\circle*{1}}
\put(125.7, 86.7){\circle*{1}}

\put(127.3, 84.3){\circle*{1}}
\put(129.0, 82.0){\circle*{1}}
\put(130.9, 80.0){\circle*{1}}
\multiput(132.9,78.0)(2.0,-2.0){3}{\circle*{1}}

\put(100,80){\line(1,0){7}}
\put(100,77){\line(1,0){9}}
\put(100.8,74){\line(1,0){9}}
\put(102.1,71){\line(1,0){7.5}}
\put(103.9,68){\line(1,0){5.5}}
\put(107,65){\line(1,0){2}}

\end{picture}

\vskip 20pt 
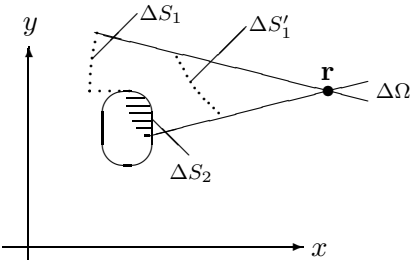
\noindent {\bf Figure 8}

\setlength{\unitlength}{0.013in} 
\begin{picture}(200,160)

\multiput(100,20)(0,75){2}{\line(0,1){35}}
\multiput(140,20)(0,75){2}{\line(0,1){35}}

\multiput(100,35)(0,5){5}{\line(1,0){40}}
\multiput(100,95)(0,5){5}{\line(1,0){40}}
\multiput(85,45)(0,3){21}{\circle*{1}}
\multiput(85,45)(3,0){6}{\circle*{1}}
\multiput(85,105)(3,0){6}{\circle*{1}}

\put(160,58.5){\vector(4,-1){1}}
\put(160,77.8){\vector(3,-1){1}}
\put(160,84.5){\vector(3,-1){1}}

\put(95,75){\line(4,-1){63}}
\put(110,95){\line(3,-1){48}}
\put(130,95){\line(3,-1){28}}

\put(130,95){\line(-1,-4){10}}
\put(120,56){\line(-3,1){25}}
\put(110,95){\line(-2,-1){15}}

\put(74,100){\makebox(10,8)[l]{\scriptsize $S$}}
\put(152,50){\makebox(10,8)[l]{\scriptsize $1$}}
\put(152,68){\makebox(10,8)[l]{\scriptsize $2$}}
\put(152,90){\makebox(10,8)[l]{\scriptsize $3$}}
\end{picture}

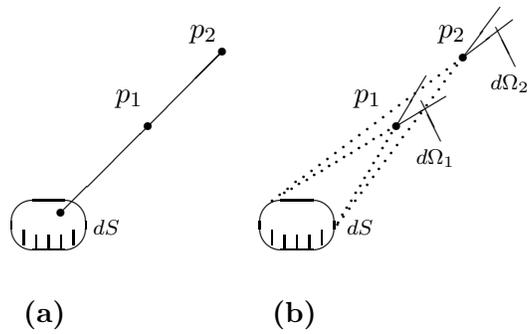
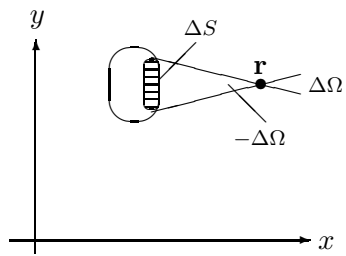
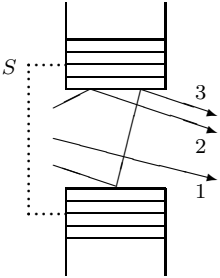
\end{document}